\documentclass[12pt]{article}
\usepackage{epsfig}

\newcommand{\AmS}{{\protect\the\textfont2
  A\kern-.1667em\lower.5ex\hbox{M}\kern-.125emS}}
\newcommand{\Dlr}{\stackrel{\leftrightarrow}{D}}
\newcommand{\Dl}{\stackrel{\leftarrow}{D}}
\newcommand{\Dr}{\stackrel{\rightarrow}{D}}

\newcommand{\Dlrsl}{\not{\hspace{-1.5mm}{\Dlr}}}
\newcommand{\pslash}{\not{\hspace{-0.08cm}p}}

\newcommand{\cswv}{c_{sw}}
\newcommand{\dst}{\displaystyle}
\newcommand{\half}{\frac{1}{2}}

\begin{document}
\date{November 5, 1997}
\title{
\vspace{-16mm} 
\rightline{\small UL-NTZ 37/97} 
\vspace{-10pt} 
\rightline{\small HUB-EP-97/75}
\vspace{-10pt}
\rightline{\small DESY 97-216}
Local bilinear operators on the lattice and their perturbative
renormalisation including $O(a)$ effects
}
\author{S.~Capitani$^1$, M.~G\"ockeler$^2$,
         R.~Horsley$^3$,  
         H.~Perlt$^4$, \\ P.~Rakow$^5$,
 	G.~Schierholz$^{1,5}$, A.~Schiller$^4$\thanks{Talk given by A.~Schiller
 	at 2nd SPIN Workshop, Zeuthen, September 1-5, 1997} 
 	\vspace{6mm}\\
{\small \it $^1$ DESY-Theory Group, Notkestra\ss e 85, D-22607 Hamburg, Germany} \\
{\small \it $^2$ Universit\"at Regensburg, Institut f\"ur Theoretische
Physik, D-93040 Regensburg, Germany} \\
{\small\it $^3$ Humboldt-Universit\"at, Institut f\"ur Physik,
Invalidenstra\ss e 110, D-10115 Berlin, Germany} \\
{\small \it $^4$ Universit\"at Leipzig, Institut f\"ur Theoretische Physik  
and NTZ,}  \\
{\small \it Augustusplatz 10-11, D-04109 Leipzig, Germany} \\
{\small \it $^5$ DESY-I\hspace{0.5mm}f\hspace{0.4mm}H Zeuthen, 
Platanenallee 6, D-15738 Zeuthen, Germany }
}

\maketitle
\vspace{-4mm}
\begin{abstract}
Some basic concepts are discussed  to  derive renormalisation
factors of local lattice  operators relevant to deep inelastic structure
functions and to other measurable quantities.
These $Z$ factors can be used to relate matrix elements measured  by lattice
techniques to their continuum counterparts.
We discuss the $O(a)$ improvement of point and one--link lattice 
quark operators.
Suitable bases of improved operators are derived. Tadpole improvement 
is applied to get more reliable perturbative
results.
\end{abstract}

\section{Introduction and some basic definitions of DIS and OPE}

In deep inelastic lepton scattering (DIS, see e.g. \cite{Manohar}) (with 4--momenta $k$ and $k'$) 
on hadron ($p$ with $p^2=M^2$)
the inclusive  
differential cross section for $e P \to e' X$  in the hadron rest frame 
is given
by 
\begin{equation}
\frac{d^3\sigma}{dx \; dy \; d\phi}= \frac{e^4}{16 \pi^2 Q^4} 
y \; l^{\mu\nu} (k,q,s_l) \; W_{\mu\nu} (p,q)_{\lambda\lambda}
\end{equation}
with the standard notations
\begin{equation} 
x=\frac{Q^2}{2 p \cdot q}\, , \ \ \ 
y=\frac{ p \cdot q}{p \cdot k}\, , 
\end{equation} 
$q=k-k'$ (with $-q^2=Q^2$) is the momentum transfer in the scattering
process,
$\phi$ the azimuthal scattering angle of the outgoing lepton,
$s_l$ (with $s_l^2=-m_l^2$) the initial lepton polarisation vector and $\lambda$
denotes the initial hadron polarisation ($\pm 1/2$ for spin 1/2 target).
All information about the cross section is contained in the leptonic 
and hadronic
tensors $l_{\mu\nu}$ and $W_{\mu\nu}$.
While $l_{\mu\nu}$ is known, the hadronic tensor  
\begin{equation}
W^{\mu\nu}(p,q)_{\lambda'\lambda}= \frac{1}{4\pi} \int d^4 x \; 
{\mathrm e}^{iq \cdot x}
\langle p,\lambda'| [ j^\mu (x), j^\nu (0) ] | p, \lambda \rangle
\end{equation}
with the electromagnetic hadronic currents $j^\mu (x)$ contains the
strong interaction effects which are not completely accessible to
perturbative QCD.

The most general hadronic tensor for polarised DIS from spin $1/2$ targets
is usually written in the form
\begin{eqnarray}
W_{\mu\nu}(p,q,s)&= & F_1\; \left( -g_{\mu\nu}+ \frac{q_\mu q_\nu}{q^2}\right) +
\frac{F_2}{p \cdot q} \; \left(p_\mu - \frac{ p \cdot q}{q^2} \; q_\mu\right)
                         \left(p_\nu - \frac{ p \cdot q}{q^2} \; q_\nu\right)
\nonumber \\
&+& i \epsilon_{\mu\nu\rho\sigma} \frac{q^\rho}{p \cdot q}
\left( (g_1+g_2) s^\sigma - g_2 \; \frac{s \cdot q}{p \cdot q} \; p^\sigma  \right)
\, .
\end{eqnarray}
Here  $F_{1,2}$ and $g_{1,2}$
are the structure functions and $s_\mu$ the polarisation vector of the nucleon
with $s^2=-M^2$.
In a parton model interpretation the structure functions $F_{1,2}$ contain 
information about the overall density of quarks (and gluons) in the nucleon and
$g_1$ probes the distribution of quarks of given helicity in a longitudinally
polarised nucleon ($Q_i$ is the quark charge, $q_\pm(x)({\overline q}_\pm(x))$ 
denotes the  distribution function of  quark (antiquark) 
with momentum fraction $x$  and helicity 
parallel/antiparallel to its parent nucleon)
\begin{eqnarray}
F_1(x,Q^2)&=& \sum_i \frac{Q_i^2}{2} 
\left(q_+(x)+q_-(x)+\overline{q}_+(x) + \overline{q}_-(x)\right) \, ,
\nonumber \\
F_2(x,Q^2)&=& \sum_i Q_i^2  x\left( q_+(x)+q_-(x)+\overline{q}_+(x) 
+ \overline{q}_-(x)\right) \, ,
  \\
g_1(x,Q^2)&=& \sum_i \frac{Q_i^2}{2} 
\left( q_+(x)-q_-(x)+\overline{q}_+(x) - \overline{q}_-(x)\right) \, .
\nonumber
\end{eqnarray}
The structure function $g_2$ has no simple interpretation in the parton model.

One can derive sum rules (moments) for the structure functions directly 
from QCD 
using the operator product expansion (OPE).
The starting point is the time--ordered product of two hadronic currents
\begin{equation}
t_{\mu\nu} = i \int d^4 z \; {\mathrm e}^{i q \cdot z} \;
T \left( j_\mu\left(z\right) j_\nu\left(0\right) \right)
\end{equation}
the matrix element of which gives the well--known Compton amplitude.
This amplitude
is related to the hadronic tensor $W_{\mu\nu}$ via the optical theorem
\begin{equation}
2 \pi W_{\mu\nu}(p,q)_{\lambda'\lambda} = 
{\mathrm {Im}} \;\langle p, \lambda' | t_{\mu\nu} | p, \lambda \rangle \, .
\end{equation}

To calculate the Compton amplitude in QCD one relies on  OPE which
allows to write the product of two local operators $O_a(z)$ and 
$O_b(0)$ for vanishing
distance $z$ as an expansion in local
operators. In the Fourier transform version of OPE one has
\begin{equation}
\lim_{q \to \infty} \int d^4z \; {\mathrm e}^{i q \cdot z} \; O_a(z) O_b(0) 
= \sum_d c_{abd}(q) O_d(0) \, .
\end{equation} 
The Wilson coefficients $c_{abd}(q)$ (in general singular at $q \to \infty$)
are independent of the matrix elements, provided $q$
is much larger than the characteristic momentum in any of the external states.
 
 The local operators in OPE for QCD are quark and gluon operators
 with arbitrary dimension $d$ and spin $n$.
 It can be shown that the contribution of any operator to 
 $l^{\mu\nu} W_{\mu\nu}$ is of the order
 \begin{equation}
 c_{\mu_1\dots\mu_n} O_{d,n}^{\mu_1\dots\mu_n} \propto x^{-n} 
 \left( \frac{Q^2}{M^2} \right)^{(2-t)/2}
\end{equation} 
 with the twist introduced as $t=d-n$.
 Therefore, the most important operators in OPE at $Q^2\to \infty$ are those 
 with twist two contributing to $F_{1,2}$ and $g_1$. 
 The structure function $g_2$ involves  twist three operators allowing a 
direct
 measurement of higher twist matrix elements.
 
For unpolarised DIS we use  as conventional basis for twist 
two quark and gluon operators 
\begin{eqnarray}
{\cal O}_{\mu_{1} \dots \mu_{n}}^{(u,d)}&=&\left(\frac{i}{2}\right)^{n-1}
\bar{\psi}^{(u,d)}\gamma_{\mu_1} \Dlr_{\mu_2}\dots \Dlr_{\mu_n} \psi^{(u,d)} -
{\mathrm{traces}} 
\, , \label{9a} \\
 {\cal O}_{\mu_{1} \dots \mu_{n}}^{(g)}&=&{i}^{n-2}
{\rm Tr} F^\alpha_{\ \mu_1}  D_{\mu_2}\dots D_{\mu_{n-1}}F_{\alpha\mu_n}  -
{\mathrm{traces}}
\label{9b} 
\end{eqnarray}
where $\psi^{(u,d)}$ denote the quark fields,  
$F_{\alpha\beta}$
is the gluon field strength tensor and the covariant derivatives $D_\mu$ and
$\Dlr_\mu=\Dr_\mu-\Dl_\mu$.
Taking into account polarisation, the following towers of
 operators have to be 
added
\begin{eqnarray}
{\cal O}_{\sigma \mu_{1} \dots \mu_{n}}^{5(u,d)}&=&\left(\frac{i}{2}\right)^{n}
\bar{\psi}^{(u,d)}\gamma_\sigma \gamma_{5} \Dlr_{\mu_1}\dots \Dlr_{\mu_n} 
\psi^{(u,d)} -
{\mathrm{traces}} 
\, , \label{10a} \\
{\cal O}_{\sigma \mu_{1} \dots \mu_{n}}^{5(g)}&=&{i}^{n-1}
{\rm Tr} \widetilde{F}^\alpha_{\ \sigma}  D_{\mu_1}\dots D_{\mu_{n-1}}
F_{\alpha\mu_n}  - {\mathrm{traces}} 
\label{10b}
\end{eqnarray}
with the dual field strength tensor $\widetilde{F}_{\alpha\beta}= \frac12
\epsilon_{\alpha\beta\gamma\delta}F^{\gamma\delta}$.

The moments of nucleon structure functions are then written at large $Q^2$
in the form
\begin{eqnarray}
2 \int_0^1 dx \ x^{n-1} F_1(x,Q^2)&=&  
 \sum_{f=u,d,g} c_{1,n}^{(f)}\Big(\frac{\mu^2}{Q^2},g(\mu)\Big) \  {v_n^{(f)}(\mu)} 
\, , \nonumber \\
\int_0^1 dx \ x^{n-2} F_2(x,Q^2)&=& 
 \sum_{f} c_{2,n}^{(f)}\Big(\frac{\mu^2}{Q^2},g(\mu)\Big) \  {v_n^{(f)}(\mu)} 
\end{eqnarray}
and 
\begin{eqnarray}
2 \int_0^1 dx \ x^{n} g_1(x,Q^2)&=&
\frac12\sum_{f=u,d,g}  e_{1,n}^{(f)}\Big(\frac{\mu^2}{Q^2},g(\mu)\Big) \ 
 {a_n^{(f)}(\mu)}
\, ,   \\ 
 2 \int_0^1 dx \ x^{n} g_2(x,Q^2)&=&
\frac12 \frac{n}{n+1}   
  \sum_{f=u,d,g}  \left( e_{2,n}^{(f)}\Big(\frac{\mu^2}{Q^2},
g(\mu)\Big) \  {d_n^{(f)}(\mu)}
 - e_{1,n}^{(f)}\Big(\frac{\mu^2}{Q^2},g(\mu)\Big) \  {a_n^{(f)}(\mu)}\right)
\, .\nonumber
\end{eqnarray}
Due to the symmetry of the structure functions under
charge conjugation only even $n$ contribute to $F_{1,2}$ and $g_{1,2}$.
In the so called quenched approximation of lattice QCD, however, also 
operators with odd $n$ are relevant.
The coefficient functions  $c_{1,n}$, $c_{2,n}$, $e_{1,n}$ and $e_{2,n}$
are calculable in QCD perturbation theory, the measured scaling violations
are usually described by their $Q^2$ evolution.

On the other hand the computation of structure functions themselves (at a given
low momentum scale $\mu$) requires nonperturbative methods ab initio.
Note that $\mu$--dependence in the matrix elements $v_n^{(f)}(\mu)$, 
$a_n^{(f)}(\mu)$ and $d_n^{(f)}(\mu)$ arising from the $\mu$--dependence
of the renormalisation constants defined below
should be cancelled by the corresponding $\mu$--dependence in the coefficient 
functions.
The matrix elements are defined from the operators as follows
\begin{eqnarray}
 \frac12 \sum_{\vec{s}} 
\langle \vec{p},\vec{s}| {{\cal O}_{\{\mu_{1} \dots \mu_{n}\}}^{(f)}}
| \vec{p},\vec{s} \rangle &=& 
 2  {v_n^{(f)}}(p_{\mu_1}\dots p_{\mu_n} -{\mathrm {traces}})
 \, , \label{13a} \\
\langle \vec{p},\vec{s}| {
{\cal O}_{\{\sigma \mu_{1} \dots \mu_{n}\}}^{5(f)}}
| \vec{p},\vec{s} \rangle &= &
  \frac{1}{n+1}   {a_n^{(f)}}(s_\sigma p_{\mu_1}\dots p_{\mu_n}+\dots -
{\mathrm{traces}})
\, , \label{13b} \\
 \langle \vec{p},\vec{s}|
 {{\cal O}_{ [ \sigma \{ \mu_1] \dots \mu_n\}}^{5(f)}}
| \vec{p},\vec{s} \rangle &=& 
  \frac{1}{n+1}   {d_n^{(f)}}\ (s_{[\sigma} p_{\mu_1]}p_{\mu_2}\dots 
p_{\mu_n}+\dots -
{\mathrm{traces}}) \, .
\label{13c}  
\end{eqnarray}
Here $\{ \dots \}$ denotes symmetrisation and $[ \dots ]$ antisymmetrisation.

The traceless and symmetric operators 
${\cal O}_{\{\mu_1 \cdots \mu_n\}}^{5(f)}$ and
${\cal O}_{\{\sigma \mu_1 \cdots \mu_n\}}^{5(f)}$  transform 
irreducibly under
the Lorentz group. The r.h.s. of eqs.~(\ref{13a}) and (\ref{13b}) are the
only traceless, symmetric tensors of maximum spin --
i.e. $n$ and $n+1$, respectively -- one can
build from a single momentum vector and the polarisation vector $s_\mu$.
Both operators have twist two. The operator
${\cal O}_{[\sigma\{\mu_1] \cdots \mu_n\}}^{5(f)}$, which is also traceless but of
mixed symmetry, transforms irreducibly as well  
and has spin $n$ and twist three.

Besides the operators (\ref{9a}-\ref{10b}) used in DIS also point quark operators
are of interest to calculate e.g. masses or decay constants: 
\begin{equation}
{\cal O}_\Gamma=
\bar{\psi}\Gamma \psi 
\label{14} 
\end{equation}
with
\begin{equation}
\Gamma= 1,\ \gamma_\mu, \gamma_5, \ \gamma_\mu\gamma_5, \ \sigma_{\mu\nu} 
, \ \sigma_{\mu\nu}\gamma_5\, .
\label{15}
\end{equation}

The matrix elements for the lowest spins are calculable on the theoretical basis
of lattice QCD using numerical simulation techniques which allow  in principle
to define the   structure functions as  physical observables
on one common theoretical basis QCD.
A lot of results for the matrix elements has 
been obtained by our QCDSF--collaboration,
part of them are discussed in the talk of 
G.~Schierholz \cite{GS} at this workshop.

\section{Action and operator improvements}

To reduce  
systematic discretisation errors    in realistic 
lattice calculations to $O(a^2)$,
an improved action, proposed by Sheikholeslami and Wohlert \cite{SW} can 
be used.  In their approach a  higher dimensional
operator is added to   the Wilson action $S_W$ which is restricted by the 
same symmetries 
as the original unimproved action ($a$ is the lattice spacing, $r$ the 
Wilson coefficient) 
\begin{equation}
S_{\rm imp} = S_W + 
     c_{sw} \,g\, i\, \left(a^4 \sum_{x,\mu\nu}\frac{a r}{4}
     \bar{\psi}(x)\sigma_{\mu\nu} F_{\mu\nu}^{\rm clover}(x) \psi(x) \right) 
     \,  .
\end{equation}
$F_{\mu\nu}^{\rm clover}$ denotes the clover leaf form of the lattice 
field strength, 
$\sigma_{\mu\nu}= (i/2)(\gamma_\mu \gamma_\nu- \gamma_\nu \gamma_\mu)$.
The constant $c_{sw}$ which gives the strength of the higher dimensional 
operator  is given in a perturbative expansion as $ c_{sw}=1 + 0.2659 g^2$
\cite{wohlert,LW}.
The coefficient can be tuned to obtain on--shell improved Green's functions.
The extra action piece adds a vertex contribution to the
 lattice Feynman rules.

In Euclidean space-time
the Lorentz group is replaced by the orthogonal group $O(4)$, which on the
lattice reduces to the hypercubic group $H(4) \subset O(4)$. Accordingly, the
lattice operators are classified by their transformation properties
under the hypercubic group and charge conjugation.
In ref.~\cite{opus4} we have identified all irreducible representations
of the operators ${\cal O}$ and ${\cal O}^5$ up to rank four.
In the (Wick rotated) operators    
the covariant
derivatives are replaced by the lattice
covariant derivatives (with the link matrix $U_{x,\mu}$)
\begin{eqnarray}
\stackrel{\rightarrow}{D}_{\mu} \psi(x) & = & \frac{1}{2a}
\left( U_{x,\mu} \psi(x+\hat{\mu}) - U^\dagger_{x-\hat{\mu},\mu}
\psi(x-\hat{\mu})\right)  
 \, , \nonumber \\
\bar{\psi}(x)\stackrel{\leftarrow}{D}_{\mu}  & = & \frac{1}{2a}
\left( \bar{\psi}(x+\hat{\mu})U^\dagger_{x,\mu} - \bar{\psi}
(x-\hat{\mu})U_{x-\hat{\mu},\mu}\right)\, . 
\label{d2}
\end{eqnarray}
 
To be consistent in an improvement program, besides of an improved action, 
the operators under discussion
have to be improved, too.
We have constructed the fundamental bases necessary to achieve full
$O(a)$ improvement of point and one--link quark operators.
This is achieved by
adding higher dimensional operators with the same 
symmetry properties (parity, charge conjugation)
as the original unimproved ones. The bases for point and one--link operators
are listed in the following:
\vspace{3mm} \\
{$\bullet S=\bar{\psi}\psi$}
\begin{equation}
\left(\bar{\psi}\psi\right)^{\rm imp} = (1+a \,b\, m)\bar{\psi}\psi -
\frac{1}{2} a c_1 \bar{\psi}\Dlrsl\psi
\, ,
\end{equation}
{$\bullet A_5=\bar{\psi}\gamma_5\psi$}
\begin{equation}
\left(\bar{\psi}\gamma_5\psi\right)^{\rm imp}  = (1+a \,b\, m)\bar{\psi}\gamma_5
\psi +
\frac{1}{2}a c_1 \partial_\mu \Big(\bar{\psi}\gamma_\mu\gamma_5\psi\Big)
\, ,
\end{equation}
{$\bullet V_\mu=\bar{\psi}\gamma_\mu\psi$}
\begin{equation}
\left(\bar{\psi}\gamma_\mu\psi\right)^{\rm imp}  =  (1+a\, b\, m)\bar{\psi}\gamma_
\mu\psi - 
\frac{1}{2}a c_1 \bar{\psi}\Dlr_\mu \psi  
 + i \half a c_2 \partial_\lambda\Big(\bar{\psi}\sigma_{\mu\lambda}\psi\Big)
\, ,
\end{equation}
{$\bullet A_\mu=\bar{\psi}\gamma_\mu\gamma_5\psi$}
\begin{equation}
\left(\bar{\psi}\gamma_\mu\gamma_5\psi\right)^{\rm imp}  = 
(1+a\, b\, m)\bar{\psi}\gamma_\mu\gamma_5\psi - 
i \half a c_1 \bar{\psi}\sigma_{\mu\lambda}\gamma_5\Dlr_\lambda \psi  
 + \half a c_2 \partial_\mu\Big(\bar{\psi}\gamma_5\psi\Big)
\, ,
\end{equation}
{$\bullet t_{\mu\nu}=\bar{\psi}\sigma_{\mu\nu}\psi$}
\begin{equation}
\left(\bar{\psi}\sigma_{\mu\nu}\psi\right)^{\rm imp} = 
(1+a \,b\, m)\bar{\psi}\sigma_{\mu\nu}\psi + 
i \half a c_1 \epsilon_{\mu\nu\lambda\tau}
 \bar{\psi}\gamma_\tau\gamma_5\Dlr_\lambda\psi 
 + i \half a c_2
              \left(\partial_\mu\Big(\bar{\psi}\gamma_\nu \psi\Big) -
                    \partial_\nu\Big(\bar{\psi}\gamma_\mu \psi\Big)\right)
\, ,
\end{equation}
{$\bullet (h_1)_{\mu\nu}=\bar{\psi}\sigma_{\mu\nu}\gamma_5\psi$ 
for the Drell-Yan process}
\begin{equation}
\left(\bar{\psi}\sigma_{\mu\nu}\gamma_5\psi\right)^{\rm imp} = 
(1+a \,b\, m)\bar{\psi}\sigma_{\mu\nu}\gamma_5\psi 
+ i \half a c_1\bar{\psi}
     \left(\gamma_\mu\Dlr_\nu -\gamma_\nu\Dlr_\mu\right)\gamma_5 \psi 
+ i \half a c_2 \epsilon_{\mu\nu\lambda\tau}\partial_\tau\Big(\bar{\psi}
\gamma_
\lambda\psi\Big)
\, ,
\end{equation}
{$\bullet {\cal O}_{\mu\nu}=\bar{\psi}\gamma_\mu \Dlr_\nu \psi$}
\begin{eqnarray}
{\cal O}_{\mu\nu}^{\rm imp,1} & = &
\left(1+ a\, b\, m\right)\bar{\psi}\gamma_\mu \Dlr_\nu \psi -
  a\, c_1^{(1)}\, g \,\bar{\psi}\sigma_{\mu\lambda} F_{\nu\lambda}^{\rm clover}
\psi  
\nonumber \\
 &  -& \frac{1}{4}\,a \,c_2 \, \bar{\psi}\left\{\Dlr_\mu,\Dlr_\nu\right\}\psi   +
 \frac{1}{2}\,a \,i\,c_3 \partial_\lambda\Big(\bar{\psi}\sigma_{\mu\lambda}
\Dlr_ \nu\psi\Big) \, , \\
{\cal O}_{\mu\nu}^{\rm imp,2} & = &
\left(1+ a\, b\, m\right)\bar{\psi}\gamma_\mu \Dlr_\nu \psi +
\frac{1}{4}\,a\,i c_1^{(2)}  
\bar{\psi}\sigma_{\mu\lambda}\left[\Dlr_\nu,\Dlr_\lambda\right]\psi 
\nonumber  \\
&  -&\frac{1}{4} \,a \,c_2  \bar{\psi}\left\{\Dlr_\mu,\Dlr_\nu\right\}\psi   +
 \frac{1}{2}\,a \,i\,c_3 \partial_\lambda\Big(\bar{\psi}\sigma_{\mu\lambda}
\Dlr_ \nu\psi\Big)
\, ,
\end{eqnarray}
{$\bullet {\cal O}_{\mu\nu}^5=\bar{\psi}\gamma_\mu \gamma_5 \Dlr_\nu \psi$}
\begin{eqnarray}
{\cal O}_{\mu\nu}^{5,\rm imp,1} & = &
\left(1+ a\, b\, m\right)\bar{\psi}\gamma_\mu\gamma_5 \Dlr_\nu \psi -
  a\, i \,c_1^{(1)}\, g \,\bar{\psi}\gamma_5 F_{\mu\nu}^{\rm clover}\psi 
\nonumber \\
&  -&\frac{1}{4}\,a \,i \,c_2 \, \bar{\psi}\sigma_{\mu\lambda}
\gamma_5\left\{\Dlr_\lambda,\Dlr_\nu\right\}\psi   +
 \frac{1}{2}\,a \,i\,c_3 \partial_\mu\Big(\bar{\psi}\gamma_5
\Dlr_\nu\psi\Big) 
\, , \label{31} \\
{\cal O}_{\mu\nu}^{5,\rm imp,2} & = &
\left(1+ a\, b\, m\right)\bar{\psi}\gamma_\mu\gamma_5 \Dlr_\nu \psi -
\frac{1}{4}\, a\, c_1^{(2)}\, g \,\bar{\psi}\gamma_5 
\left[\Dlr_\mu,\Dlr_\nu\right]\psi \nonumber \\
&  -&\frac{1}{4}\,a \,i \,c_2 \, \bar{\psi}\sigma_{\mu\lambda}\gamma_5
\left\{\Dlr_\lambda,\Dlr_\nu\right\}\psi   +
 \frac{1}{2}\,a \,i\,c_3 \partial_\mu\Big(\bar{\psi}\gamma_5
\Dlr_\nu\psi\Big)
\label{32}
\, .
\end{eqnarray}

Due to
$
\left[\Dlr_\mu,\Dlr_\nu\right]^{\rm lattice} 
= 4 \,i \,g \, F_{\mu\nu}^{\rm clover} + O(a^2)
 $
two prescriptions for the improved
one--link lattice operators are possible.

Inserting  the improved operators  into forward matrix elements the
surface term $\partial_\mu\Big(\bar{\psi}{\cal O}\psi\Big)$
 (where ${\cal O}$ is
any operator) vanishes due to momentum conservation.
Using the equation of motion it is possible for each 
improved operator to eliminate 
one base operator.
For the coefficients  $c_i=1+O(g^2)$  
the perturbative expansion  is not known, 
$c_i$ and $b$ are different quantities for every operator considered.

\section{Renormalisation}

\subsection{Renormalisation conditions}

In order to relate the matrix elements computed on the lattice to continuum 
matrix
elements the so called $Z$ factors have to be calculated. A 
consistent way would be
to do this also nonperturbatively, e.g. on the lattice\cite{mart1}. 
Here we present    one--loop
perturbative calculations (using totally anticommuting $\gamma_5$) which
can be used as a first step to control the nonperturbative result.
 We  present  the $Z$ factors with coefficients
$c_i$ and $c_{sw}$ kept arbitrary. This allows to define the perturbative
contributions of the various terms and their relative magnitudes. Moreover,
this will allow to implement tadpole improved perturbation theory.

There are several possibilities to determine renormalised quantities which
can be compared to data obtained in experiments. We use the
projection onto the tree structure (cf. e.g. \cite{qcdsf5}). 
The finite quark operators renormalised at finite scale $\mu$ with their
multiplicative renormalisation factors are given 
by the relations
 \begin{eqnarray}
  {\cal O}(\mu) &=& Z_{{\cal O}} \; {\cal O}(a) \, , \nonumber \\
  \langle q(p)|{\cal O}(\mu) |q(p) \rangle &=& 
                Z_{{\cal O}} Z_\psi^{-1} \; \langle q(p)|{\cal O}(a) 
                    |q(p) \rangle  \, , \nonumber \\
               \langle q(p)|{\cal O}(\mu) |q(p) \rangle 
                 \Big|_{p^2=\mu^2} &=&
               \langle q(p)|{\cal O}(a) |q(p) \rangle 
                 \Big|^{\rm tree}_{p^2=\mu^2} \, . 
               \label{renscheme1}
  \end{eqnarray}
$\langle q(p)|{\cal O}(a) |q(p) \rangle$
is the amputated Green's function.
$Z_\psi$ is the wave function renormalisation factor determined either via the
quark propagator 
or via the conserved vector current\cite{mart1}
\[
Z_\psi=\frac{1}{48}{\rm tr}\Big(\Lambda_{V_\mu^c}(pa)\Big)\Big|_{p^2=\mu^2},
\] 
where $\Lambda_{V_\mu^c}$ is the amputated Green's function of the conserved 
vector current. The definition (\ref{renscheme1}) corresponds to the momentum
subtraction scheme.
  
\subsection{About the calculation and program code} 

The calculation basically amounts to the computation of
integrals of the form
\begin{equation}
{\cal I}_{\mu_1\cdots\mu_n}(a,p) = \int \frac{\mbox{d}^4 k}{(2\pi)^4}
{\cal K}_{\mu_1\cdots\mu_n}(a,p,k) \, ,
\label{34}
\end{equation}
where ${\cal K}$ contains lattice quark and gluon propagators and $\sin, \cos$
of lattice momenta $p_\mu$ and $k_\mu$,
the integration is over the first Brillouin zone 
$-\pi/a \leq k_\mu < \pi/a$.
The calculation of the loop integrals ${\cal I}$ is performed in two
parts. We decompose (\ref{34}) \cite{Kawai,qcdsf5}
\begin{equation}
{\cal I} = \tilde{\cal I} + ({\cal I} - \tilde{\cal I}) \, , 
\label{split}
\end{equation}
where
\begin{equation}
\tilde{\cal I}(a,p)   =  
        \sum_{n=0}^N \frac{p_{\alpha_1} \dots p_{\alpha_n}}{n!} 
\frac{\partial^n}{\partial p_{\alpha_1} \dots \partial p_{\alpha_n}}
{\cal I}(a,p)\Big|_{p=0}  
\label{taylor}
\end{equation}
and the order of the expansion $N$ is
determined by the degree  of ultraviolet (UV)
 divergences of ${\cal I}(a,P)$ in the limit $a \to 0$. Therefore,  
${\cal I} - \tilde{{\cal I}}$ is rendered UV finite and is
computed in the continuum. 
The Taylor expansion of the 
lattice integral (the first term) will in general 
create an infrared (IR) divergence.
To regularise the integrals 
dimensional regularisation is used. The IR poles of $\tilde{{\cal I}}$  cancel  those of 
${\cal I} - \tilde{{\cal I}}$.
UV divergent contributions ($\propto 1/a^n$) 
of the lattice integrals will cancel
in the operator representations  which we are interested in.

Let us summarise some of the basic features of the developed program code.
 \begin{itemize}
\item The program package  written in {\sl Mathematica}.
\item Symbolic Feynman rules used as input, the
   program computes one--loop forward matrix elements of bilinear
   quark and gluon operators on a hyper-cubic lattice including $O(a)$
   contributions
   and in   the continuum in symbolic form.
\item {Special features of the program:}
  \begin{itemize}
  \item Dimensional regularisation
  \item Symmetrisation tables are used to accelerate the momentum
        integration over the Brillouin zone
  \item General index handling in the complicated case of hyper-cubic
        ${\rm H}(4)$ symmetry (non-Lorentz covariant structures)
  \item Algebraic isolation of the infrared poles which leads to an exact
        cancellation of the divergences
  \item Results  given with  general index structure what allows an easy
       generation 
        of all group representations 
  \item All {\sl finite} lattice integrals represented by symbols  which
       are accurately calculated numerically
  \end{itemize}
\end{itemize}
A part of the results has been checked in a completely independent
 calculation based on a code
in {\sl Form}. 

\subsection{Quark self energy to order $O(a)$}

First we discuss the one--loop self energy for quarks  
which  contributes to almost all matrix elements of the operators discussed
below.
To transform between various
renormalisation schemes both   lattice and continuum contributions are listed.

The bare fermion propagator is given by
\begin{equation}
 S^{-1} = i \pslash + m + a r p^2 /2 - \Sigma^{\rm latt}
\label{sinv}
\end{equation} 
where we present the self energy in the standard form
\begin{eqnarray}
\Sigma^{\rm latt} & = & \frac{g^2}{16\pi^2}C_F \left(
 i\pslash \Sigma_1^{\rm latt} + m \Sigma_2^{\rm latt} 
 + a r \left( p^2 \Sigma_3  +  m \, i \pslash \Sigma_4 +   m^2 \Sigma_5 \right)
 + O(a^2) \right) \, , \\
\Sigma^{\rm cont} & = & \frac{g^2}{16\pi^2}C_F 
    \left(i\pslash \Sigma_1^{\rm cont} + m \Sigma_2^{\rm cont}\right)
    \label{sigdef} 
\end{eqnarray}
with $C_F=4/3$ for SU(3). 
The bare mass $m$ is defined by 
\begin{equation}
 m a \equiv \frac{1}{2 \kappa} - \frac{1}{2 \kappa_c} 
  = \frac{1}{2 \kappa} - 4 - \frac{g^2}{16\pi^2} C_F \Sigma_0  \, . 
\end{equation}
The perturbative calculation
is performed by expanding the massive fermion propagator in the mass 
parameter up to
order ${\cal O}(m^2)$ for $m^2 \ll p^2$.
For $r=1$ we obtain in  covariant gauge
\begin{eqnarray*}
\Sigma_0 &=& -51.4347 + 13.7331 \, c_{sw} + 5.7151 \, c_{sw}^2,\\
\Sigma_1^{\rm latt} &=&  + 16.6444  - 2.2489 \, c_{sw}   - 1.3973 \, c_{sw}^2
+  \alpha   L(ap) 
-\alpha \, ,\\
\Sigma_1^{\rm cont} &=&-  \alpha   K(\epsilon,p/\mu)
-\alpha \, ,\\
\Sigma_2^{\rm latt} &=& + 11.0680 - 9.9868 \, c_{sw}  - 0.0169 \, c_{sw}^2
+(3+ \alpha ) L(ap)
-2 \, \alpha \, ,  \\              
\Sigma_2^{\rm cont} &=& 
-(3+ \alpha) K(\epsilon,p/\mu)
-4 - 2 \, \alpha \, , \\ 
\Sigma_3 &=& 
 + 7.13 89+ 0.4857 \, c_{sw} - 0.0817 \, c_{sw}^2 - 0.0719 \, \alpha + (-3 + 3 \, c_{sw} 
 + 2 \, \alpha ) \frac{1}{2} L(ap)\, ,  \\ 
\Sigma_4 &=& 
 -6.3466 - 1.4850 \, c_{sw} + 1.2860 \, c_{sw}^2 + 0.1437 \, \alpha + (-3 - 3 \, c_{sw} 
 - 2 \, \alpha ) \frac{1}{2} L(ap) \, ,  \\ 
\Sigma_5 &=& -14.9857 +16.9857 \, c_{sw} -1.5234 \,  c_{sw}^2 
+ 2.0719 \, \alpha +(-9 + 6 \, c_{sw} - \alpha ) \frac{1}{2} L(ap)     
\end{eqnarray*}
with
\begin{equation}
L(x)   =    \gamma_E - F_0 +  \ln x^2 \, , \ \ \
K(\epsilon,x)   =   \frac{1}{\epsilon}  -\gamma_E +\ln 4\pi  - \ln x^2  \, .
\end{equation} 
$\alpha$ is the gauge parameter ($\alpha=1(0)$ for Feynman (Landau) gauge),
$F_0 = 4.369225$, $\gamma_E=0.5772\dots$

\subsection{Operator renormalisation}
The renormalisation factors $Z_{\cal O}$ in the momentum subtraction scheme
 are given in the form
\begin{equation}
Z_{\cal O}(a\mu,g) = 1-\frac{g^2}{16\pi^2} C_F \left(\gamma_{\cal O} \ln (a\mu) +B_{\cal O}\right)
+O(g^4),
\label{ZO}
\end{equation} 
where $\gamma_{\cal O}$ is the anomalous dimension and $B_{\cal O}$ the finite part
of $Z_{\cal O}$. As can be seen from (\ref{renscheme1}) 
$B_{\cal O}$ receives contributions from  
the one--loop amputated Green's function  and the self energy 
diagrams (wave function renormalisation) 
\begin{equation}
B_{\cal O} = B_{\cal O}^{\rm amputated} + B_{\cal O}^{\rm self}
\end{equation} 
with 
\[
B^{\rm self}_{\cal O}  = \Sigma^{\rm latt}_1 -\alpha   L(ap) \, .
\]

Since the Wilson coefficients are
usually computed in the MS or $\overline{{\rm MS}}$ scheme,
one has to present  the renormalisation constants in these schemes, too.
The transformations between the different schemes are \cite{mart1,qcdsf5}
\begin{equation}
B_{\cal O}^{\rm MS} = B_{\cal O} - B_{\cal O}^{\rm cont} \, , \ \ \
 B_{\cal O}^{\overline{\rm MS}} = B_{\cal O}^{\rm MS}   +
\frac{\gamma_{\cal O}}{2}(\gamma_E - \ln 4\pi) \, ,
\label{trans2}
\end{equation}
where $\gamma_{\cal O}$ and $B_{\cal O}^{\rm cont}$  are given in
Table~\ref{tab:1}. 
\begin{table}[!thb]
\label{anomdimtab}
\vspace{0.5cm}
\begin{center}
\begin{tabular}{||c|c|c||}
\hline
 ${\cal O}$        &   $\gamma_{\cal O}$ &  $B_{\cal O}^{\rm cont}$   \\
\hline
 $1$, $\gamma_5$   & $-6 $               & $4 +
    \frac{\dst\gamma_{\cal O}}{\dst 2} \left(\gamma_E - \ln 4\pi\right) + \alpha$ \\
[0.7ex]
  $\gamma_\mu$, $\gamma_\mu\gamma_5$    & $0$ & $0$\\
[0.7ex]
$\sigma_{\mu \nu} \gamma_5 $        & $ 2 $ & $ 
\frac{\dst\gamma_{\cal O}}{\dst 2} \left(\gamma_E - \ln 4\pi\right) - \alpha$   \\
[0.7ex]
${\cal O}_{\{ 1 4 \}}$, ${\cal O}_{\{ 1 4 \}}^5$ &  $\frac{16}{3}$   &   
$-\frac{40}{9} + \frac{\dst\gamma_{\cal O}}{\dst 2} 
\left(\gamma_E - \ln 4\pi \right) + 1- \alpha$ \\
[0.7ex]
\hline
\end{tabular}
\caption{Anomalous dimensions and finite contributions of continuum integrals
to $B_{\cal O}$}
\label{tab:1}
 \end{center}
\end{table}
\subsection{Examples for renormalisation of   operators $A_\mu$ 
and ${\cal O}_{\mu\nu}^5$}
\vspace{-2mm}   
The  matrix element  of point quark operator $A_\mu$
to order $O(g^2)$   up to  $O(a)$  is given by the form 
\begin{equation}
\langle {A_\mu} \rangle \Big|_{g^2} =  
\frac{g^2}{16\pi^2}C_F \left(\langle{A_\mu}^{(0)}\rangle
 + a\, \langle {A_\mu}^{(1)}\rangle\right) 
\end{equation}
As result for the amputated Green's functions we find
\begin{eqnarray}
\langle A_\mu^{(0)} \rangle &=&  
  \gamma_\mu\gamma_5 \Big(
   -0.8481 
   + 2.4967 \, c_{sw}  
    -   0.8541 \, c_{sw}^2   
   - c_1 \, (19.3723 - 10.3167  \, c_{sw}  + 0.8846  \, c_{sw}^2 )
             \nonumber \\   
 & +& \alpha  -   \alpha L(ap) \,    \Big)  
   -2 \, \alpha \frac{\pslash\gamma_5 p_\mu}{p^2}
   = \gamma_\mu\gamma_5 \Big(B_{\gamma_\mu\gamma_5}^{\mathrm {amputated}}
   - \alpha L(ap)\Big) 
   -2 \, \alpha \frac{\pslash\gamma_5 p_\mu}{p^2}
    \, , \\
\langle A_\mu^{(1)} \rangle &=&   
 \frac{i}{2}(\pslash\gamma_\mu\gamma_5 + \gamma_\mu\gamma_5\pslash)\Big)\Big(
0.6760  + 4.7905 \, c_1  - 
  1.7181 \, c_{sw} + 0.5430 \, c_1 c_{sw} 
  \nonumber \\  
 &   + & 0.1302 \, c_{sw}^2 + 
  0.0537 \, c_1 c_{sw}^2 
  +\alpha (0.8563-4.0583 \, c_1)  
   +\alpha (1+ c_1)L(ap)
\, ,\\  
\langle A_\mu^{(0) {\rm cont}} \rangle &=&  \gamma_\mu\gamma_5 \,
\alpha (1+K(\epsilon,p/\mu)\,)-2 \, \alpha \frac{\pslash\gamma_5 p_\mu}{p^2} \, .
\end{eqnarray}
Note that $O(a)$ terms do not contribute to the tree--level structure, what
is valid in general.
Therefore,
no dangerous $O(g^2 a)$ or $O(g^2 a \log a)$ terms are present 
for massless quarks in
$Z$ factors defined in this momentum subtraction scheme.
The finite contribution to the $Z$ factor in ${\overline {\rm MS}}$ is then found as  
\begin{equation}
 B^{\overline{\mathrm MS}}_{\gamma_\mu\gamma_5}   
 =   15.7963 + 0.2478 \, \cswv  - 2.2514 \, \cswv^2
    - c_1 \left( 19.3723 - 10.3167 \, \cswv+ 0.8846 \, \cswv^2 \right) \, .
\end{equation}

The calculation in the case of link operators including improved action and
operators is much more cumbersome.  The $O(a)$ 
contributions are still not known yet, however, they will also not
contribute to the tree level structure. The $Z$ factors have to
be calculated in a special representation. Choosing the representation
$O^5_{\{14\}}$,   only terms $\propto c_2$ contribute, and the two prescriptions
(\ref{31}-\ref{32}) coincide
\begin{equation}
B^{\overline{\rm {MS}}}_{\cal O} = 
-4.0988  - 1.3593 \,\cswv - 1.8926 \, \cswv^2  -c_2   \left( 27.5719 - 
16.1193 \, \cswv  + 0.7570 \, \cswv^2 \right) \, . 
\end{equation}
\vspace{-10mm}  
\section{Tadpole improvement}
\vspace{-2mm} 
It is known that 
many results from (naive) lattice perturbation theory are
 in poor agreement with their numerically determined counterparts.
One of the reasons is that tadpoles (lattice artifacts) spoil the expansion.
Lepage and Mackenzie\cite{lepage} proposed a rearrangement in the perturbative
series in order to get rid of the numerically large tadpole contributions.
These contributions are included in an essentially nonperturbative way using
e.g. the measured value of the plaquette.
 
In lattice theory tadpole corrections renormalise the link operator so that
the  vev is considerably smaller than one. As a recipe ones
scales the link variables with $u_0(g^2)$ measured in Monte Carlo
\begin{equation}
u_0 = \langle \frac{1}{3} {\rm Tr}U_{\mathrm {Plaq}} \rangle^\frac{1}{4} \ .
\end{equation}
This leads to the
consequences   ($g^{\star 2}$ renormalised at some physical scale) of rescaling
the variables
\begin{equation}
g^2 \to  g^{\star 2}=u_0^{-4} \, g^2\, ,
\ \ 
c_{sw}  \to  c_{sw}^\star = u_0^3 \, c_{sw}\, ,
\ \
c_i  \to  c_i^\star = u_0^{n_i-n_D} \, c_i \, .
\end{equation}
Here 
$n_i$ denotes the number of covariant derivatives in the higher dimensional 
operator part (proportional to $c_i$ )
and 
$n_D$ the number of covariant derivatives (links) of the original operator. 

In that case one obtains a  
tadpole improved perturbative expansion of $Z$
\begin{equation}
Z_{\cal O}  \equiv \left(\frac{u_0}{u_0}\right)^{n_D-1}\, Z_{\cal O} 
                 =    u_0^{1-n_D}\left(1-\frac{g^{\star 2}}{16\pi^2}C_F
            \Big(\gamma_{\cal O} \ln (a\mu) + B^{\star}_{\cal O}\Big)\right)
                      +
                                     O(g^{\star 4})   
             =  Z^\star_{\cal O} + O(g^{\star 4})
\end{equation}
with the improved perturbative expansion
\begin{equation}
u_0 \approx 1 - \frac{g^{\star 2}}{16\pi^2}C_F \pi^2
\end{equation}
which implies
\begin{equation}
B^{\star}_{\cal O} =B_{\cal O}(c_{sw}^\star,c_i^\star) +(n_D -1) \pi^2 \ .
\end{equation}
For the representation $O^5_{\{14\}}$   we get  
\begin{equation}
B^{\star\overline{\rm {MS}}}_{{\cal O}}  =  0.3456 - 1.3593 
\, u_0^3\cswv - 1.8926\,u_0^6 \cswv^2  - c_2 \left(
   27.5719 \,u_0  
- 16.1193\,u_0^4 \cswv  
+ 0.7570 \,u_0^7 \cswv^2 \right) 
\end{equation}

\begin{figure}[!htb]
  \vspace{-5mm}
  \begin{center}
     \leavevmode
     \epsfig{file=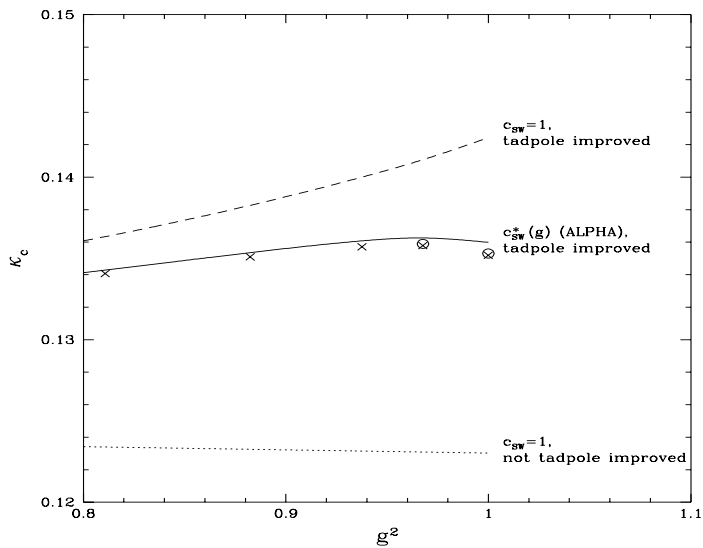,height=100mm,width=100mm}
  \end{center}
  \vspace{-30mm}
  \caption{ Dependence of critical hopping parameter $\kappa_c$ on $c_{sw}$}
  \label{fig:2}     
\end{figure}

To demonstrate the influence of tadpole improvement and addition 
of higher dimensional
operators we show in Fig.~\ref{fig:2}  \cite{ed97}    
 the dependence of the critical hopping parameter $\kappa_c$ on $\cswv$ and the gauge coupling.
The nonperturbative determination of $\cswv$ is taken from the 
ALPHA--collaboration \cite{alpha}. The tadpole improved perturbation theory
describes the Monte Carlo data (crosses, circles) significantly better, however a 
nonperturbatively
determined improvement coefficient $\cswv$ is  favoured.

\begin{figure}[!htb]
\unitlength1mm
\begin{picture}(180,80)
\put(0,0) {\epsfig{file=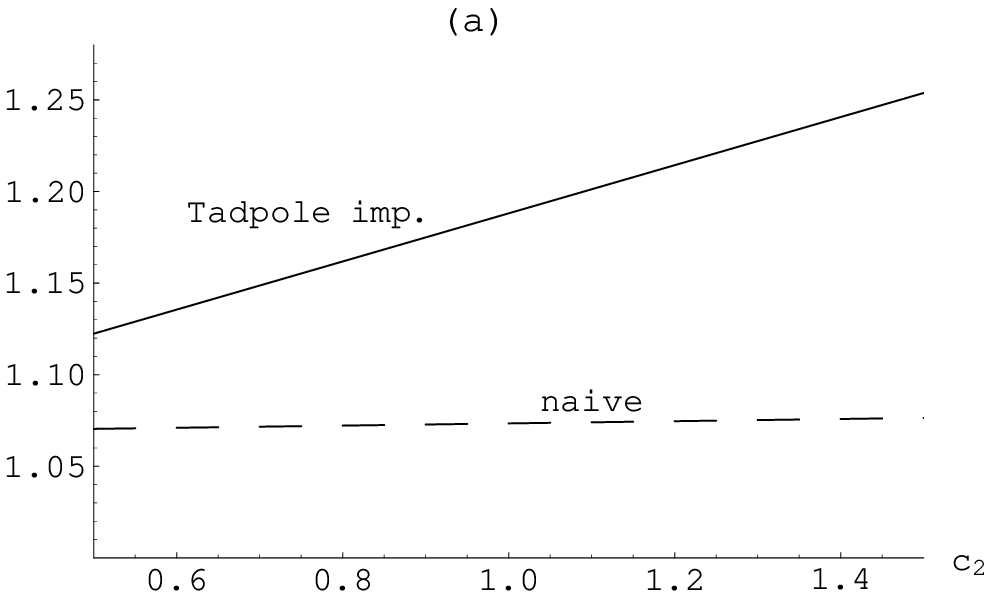,height=100mm,width=80mm}}
\put(80,0) {\epsfig{file=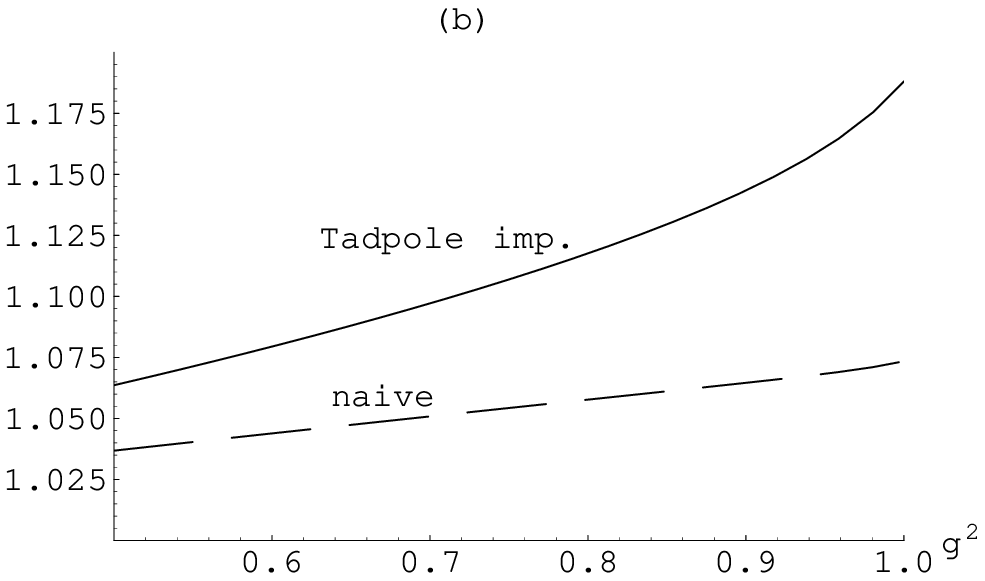,height=100mm,width=85mm}}
\end{picture}
\vspace{-25mm}
\caption{
  One-loop  $Z_{\cal O}^{\overline {MS}} $ ($O_{\{14\}}^5$ representation) 
   (a) vs. $c_2$ ($g^2=1$)  and (b) vs. $g^2$  ($c_2=1$)
   in naive and tadpole improved perturbation theory, 
   $\cswv(g^2)$ is taken from the ALPHA--collaboration}
  \label{fig:3}
\end{figure}

 In Figs.~\ref{fig:3} a and b  
perturbative predictions in $\overline {MS}$ scheme without and with tadpole 
improvement are shown 
for the renormalisation factor of the operator $O_{\{14\} }^5$  
  as function of $c_2$ and $g^2$.
Note the significant difference in the predictions.
  
\section{Summary}

1. We have developed  
an algebraic computer package   to perform
one--loop calculations in lattice  QCD perturbation theory based on 
 { \sl Mathematica}. Part of the results have been checked by a completely 
 independent code in {\sl Form}.
\\
2.
The fundamental bases are constructed which are necessary  to remove
completely $O(a)$ effects for all bilinear operators up to spin 2.
\\
3. The $Z$ factors are calculated in one--loop  for arbitrary coefficients of the 
counterterms to the operators and to the action.
\\
4. The contributions including $O(a)$ have been calculated for operators
without derivatives.
\\
5. It is planned to determine all renormalisation constants nonperturbatively.
 
Finally, Table~\ref{tab:2} 
\begin{table}[!htb]
{\small{  \begin{center}
    \leavevmode
     \begin{tabular}{|c|c|c|c|c|}
      \hline
                 &         &              &                 &      \\[-2.5ex]
      ${\cal{O}}$& $S_{W}$ & $S_{imp}$    & $S_{imp}$       & $S_{imp}$ \\
                 &         &$c_{sw},c_i=1$&$c_{sw},c_i\ne 1$&$O(a)$ \\[0.5ex]
      \hline
                 &         &              &                 &      \\[-2.5ex]
      $\bar{\psi} \psi$  & \cite{mart2} & \cite{borelli}  
                            & \cite{qcdsf4}   &  \cite{heatli},\cite{ed97}  \\
      $\bar{\psi} \gamma_5 \psi$  & \cite{mart2}  & \cite{borelli}  
                            & \cite{qcdsf4}   & \cite{heatli},\cite{ed97}   \\
      $\bar{\psi} \gamma_\sigma \psi$  &\cite{mart2}  & \cite{borelli}  
                            & \cite{qcdsf4}  & \cite{heatli},\cite{ed97}    \\
      $\bar{\psi} \gamma_\sigma \gamma_5 \psi$  &\cite{mart2}  & \cite{borelli}  
                            & \cite{qcdsf4}  &  \cite{heatli},\cite{ed97}   \\
      $\bar{\psi} \sigma_{\sigma\tau} \psi$  &\cite{mart2}  & \cite{borelli}  
                            & \cite{qcdsf4}    & \cite{heatli},\cite{ed97}  \\
                            [0.5ex]
      \hline
                 &         &              &                 &      \\[-2.5ex]
   $\bar{\psi} \gamma_\sigma \Dlr_\mu \psi$  & \cite{capi1},\cite{qcdsf5},\cite{MIT} 
                           & \cite{capi1}         & \cite{ed97}   & in preparation   \\
   $\bar{\psi} \gamma_\sigma \gamma_5 \Dlr_\mu \psi$  &\cite{MIT} &  \cite{ed97}     
                           & \cite{ed97}   & in preparation  \\
   $\bar{\psi} \gamma_\sigma \Dlr_\mu \Dlr_\nu \psi$  & \cite{capi2},\cite{qcdsf5},\cite{MIT} 
                           & \cite{capi2}  &  &   \\
   $\bar{\psi} \gamma_\sigma \gamma_5 \Dlr_\mu \Dlr_\nu \psi$  & \cite{qcdsf5},\cite{MIT} 
                           &  &  &   \\
   $\bar{\psi} \gamma_\sigma \Dlr_\mu \Dlr_\nu \Dlr_\rho\psi$  & \cite{qcdsf5},\cite{MIT} 
                           &  &  &   \\[0.5ex]
    \hline
                 &         &              &                 &      \\[-2.5ex]
  $F_{\mu\rho} F_{\rho\nu}$ & \cite{CMP},\cite{capi1},\cite{bi94},\cite{qcdsf4}
     & & & \\[0.5ex]
    \hline
    \end{tabular}
    \caption{Overview on published works on renormalisation factors of lattice bilinear 
             operators}  
  \label{tab:2}
  \end{center}
  }}
\end{table}
\vspace{-5mm}
gives an overview over the
calculated renormalisation factors of lattice bilinear operators.
We would like to mention that in \cite{LW} and \cite{SimtW} the coefficients
$b_{PS,V,A}$ and $c_i$ (for another basis of local point quark operators) 
have been obtained using the Schr\"odinger functional approach.

\end{document}